\documentclass[final,1p,times]{elsarticle}

\usepackage{amssymb}
\usepackage{subcaption}
\usepackage{lineno}
\usepackage{siunitx}
\usepackage{graphicx}
\usepackage{caption}
\usepackage{subcaption}
\usepackage{xcolor}
\usepackage{comment}
\usepackage{gensymb}
\usepackage{xcolor} 
\usepackage{hyperref}
\hypersetup{
    colorlinks=true,       
    linkcolor=blue,        
    citecolor=blue,        
    urlcolor=blue          
}
\usepackage{url}
\setcitestyle{comma}
\setcitestyle{square}

\journal{Nuclear Physics B}

\begin{document}

\begin{frontmatter}



\title{Characterization of BNL’s \SI{20}{\micro\meter} and \SI{30}{\micro\meter} thick LGADs for the Electron Ion Collider}

\author[a,b]{F. Zareef \thanks{Corresponding author, Fasih.Zareef@ifj.edu.pl}}          
\author[a]{J. J. Chwastowski}
\author[b]{M. Pycior}
\author[b]{I. Ahmad}
\author[a]{W. Fryt}
\author[c]{G. Giacomini}
\author[a]{S. Kusyk}
\author[b]{A. Obłąkowska-Mucha}
\author[a]{P. Płatek} 
\author[a]{T. Polański}
\author[a]{P. Rzeźnik} 
\author[a]{J. Swakoń} 
\author[b]{T. Szumlak}
\author[d]{A. Tricoli}
\author[a]{D. Wróbel}

\affiliation[a]{organization={The Henryk Niewodniczanski Institute of Nuclear Physics},
            addressline={Radzikowskiego 152}, 
            city={Krakow},
            postcode={31-342}, 
            country={Poland}}
\affiliation[b]{organization={AGH University of Krakow},
            addressline={Al. Adama Mickiewicza 30}, 
            city={Krakow},
            postcode={30-059}, 
            country={Poland}}

\affiliation[c]{organization={Instrumentation Department, Brookhaven National Laboratory},
            city={Upton, NY},
            postcode={11973}, 
            country={USA}}
\affiliation[d]{organization={Physics Department, Brookhaven National Laboratory},
            city={Upton, NY},
            postcode={11973}, 
            country={USA}}

\begin{abstract}

The 20 $\mu$m and 30 $\mu$m thick Low-Gain Avalanche Detectors (LGADs) produced at Brookhaven National Laboratory (BNL) were evaluated. Detailed electrical measurements indicate well-formed and stable gain layers with consistently low leakage currents. Transient Current Technique (TCT) was used to evaluate the gain. Fast-laser timing studies demonstrate time resolutions at the level of a few tens of picoseconds. 
To assess operational longevity under realistic conditions for Electron–Ion Collider sub-detector systems, devices were irradiated with 30 MeV and 60 MeV protons to fluences of 10$^{13}$ - 10$^{14}$ $n_{eq}/cm^2$, enabling a focused study of radiation-induced effects on gain stability and timing performance.
\end{abstract}

\begin{keyword}
Low Gain Avalanche Diodes (LGADs) \sep silicon detectors \sep particle trackers\sep timing detectors\sep 4D tracking  \sep radiation damage.


\end{keyword}

\end{frontmatter}


\section{Introduction}
\label{sec:intro}

The next generation of high-energy physics (HEP) experiments relies on detector systems capable of timing resolutions of O(10 ps), together with position resolutions of O(10 $\mu$m), to enable four-dimensional (4D) tracking in which both the spatial and temporal coordinates of charged particles are reconstructed \cite{Sadrozinski_2018, SADROZINSKI201618}. 
Low-Gain Avalanche Detectors (LGADs) have emerged as one of the leading silicon technologies for precision timing applications \cite{pellegrini2014technology}. LGADs incorporate a highly doped multiplication layer that generates moderate internal gain through controlled avalanche multiplication, producing large and fast signals while maintaining the advantages of silicon-based tracking detectors. 
Recent developments in thin-LGADs \cite{sola2025thin}, AC-LGADs \cite{giacomini2019fabrication}, inverse LGADs (iLGADs) \cite{curras2020inverse} and trench‑isolated LGADs (TI-LGADs) \cite{paternoster2020trench, Zareef_2025, zareef2026depth} have expanded their use in high‑rate environments where both spatial and temporal precision are required.

The Electron–Ion Collider (EIC), planned to begin operation in the early 2030s, will deliver luminosities between 10$^{33}$ and 10$^{34}$ cm$^{-2}$s$^{-1}$ and will operate with high event rates and dense forward particle fluxes \cite{yano2025physics}, however, the event rates and particle fluxes are at EIC are not very high compared to the LHC, HL-LHC or RHIC. The AC-LGADs have been identified as a leading candidate technology for EIC far-forward instrumentation, barrel-time of flight (TOF) (strips) and an endcap-TOF (pixels), where they can provide the precise timing information needed for particle identification and 4D tracking. In this work, we study standard (DC-coupled) LGADs, which share the same gain layer and thus the same bulk gain and timing mechanism as AC-LGADs. We note that obtained results are specific to standard LGADs and may not directly apply to AC-LGADs. Extending this study to AC-LGADs is left for future work.
The cumulative radiation exposure expected for EIC far-forward detectors is of the order of 10$^{13}$ 1 MeV n$_{\rm eq}$/cm$^2$ over ten years of operation \cite{STAGE2026171012}. 
This is significantly lower than the (1–3)×10$^{15}$ n$_{\rm eq}$/cm$^2$ expected for the timing layers of the HL-LHC experiments, including the ATLAS-HGTD \cite{mazza2019high} and CMS-ETL \cite{ferrero2022cms}. 

Although LGADs have shown promising radiation tolerance, their gain layer degrades under proton and neutron irradiation, leading to higher leakage currents, reduced signal amplitude, and degraded timing resolution. 
Understanding this degradation in thin sensors is essential for designing timing layers that can operate reliably in the EIC environment. 
In particular, thin-LGADs offer a balance between fast charge collection and mechanical robustness, but their behaviour under irradiation with low and medium‑energy protons remains insufficiently characterised. However, radiation effects of different proton energies on LGADs are summarized here \cite{KRAUS2026171623}. 

This work presents a systematic study of LGADs with active thicknesses of \SI{20}{\micro\meter} and \SI{30}{\micro\meter}, before and after irradiation with 30 MeV and 60 MeV proton beams with a fluence range of 10$^{13}$ - 10$^{14}$ 1 MeV n$_{\rm eq}$/cm$^2$. Single Event Burnout (SEB) is a known failure mechanism in LGADs under heavy-ion irradiation, and is observed for the same kind of devices at BNL \cite{TISHELMANCHARNY2027172107}. However, for the low-energy protons used in this work, the deposited charge and resulting field perturbation are insufficient to trigger SEB, and it was not observed. 
The goal is to quantify changes in gain, charge collection, and timing performance, and to identify sensor thickness and irradiation energy dependencies that influence long‑term operation. 
The results provide guidance for the design and qualification of LGAD‑based timing layers for the EIC and contribute to the broader understanding of LGAD performance in high‑radiation environments.

\section{Devices under test}

\begin{figure}[h]
    \centering
    \includegraphics[width=1\textwidth]{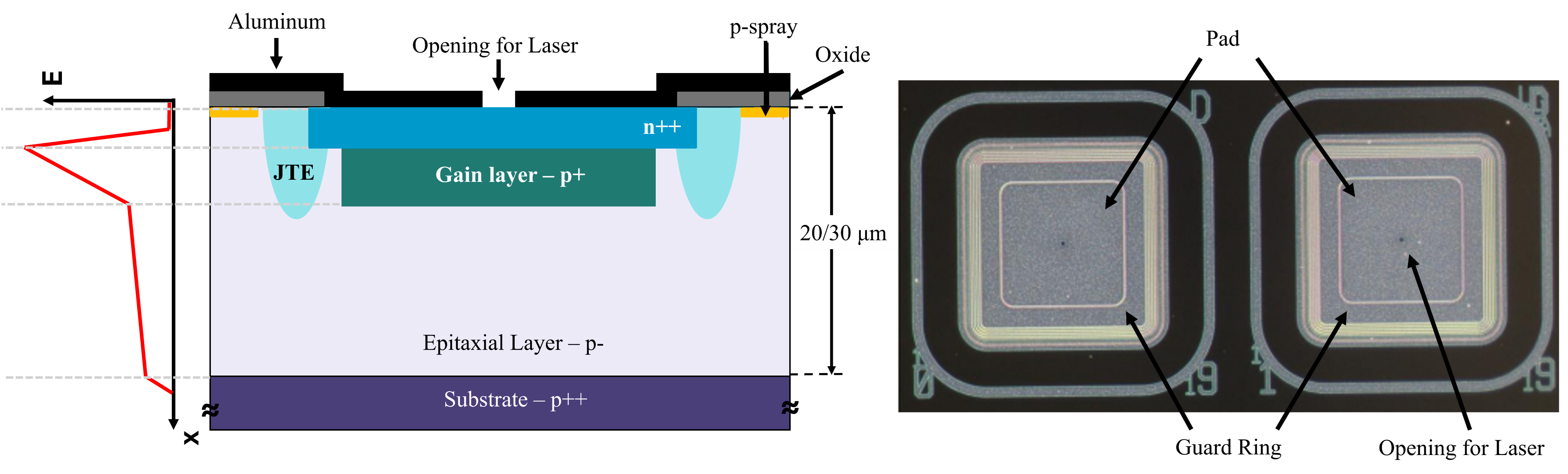}
    \caption{(Left) Schematic diagram of a standard LGAD (not to scale). (Right) Photographs of the LGAD and PIN sensors studied in this work.}
    \label{fig:schematics}
\end{figure}

The samples investigated in this study consist of LGAD and PIN sensors fabricated by BNL. 
These sensor types are identical in design except for the presence of a highly doped $p^+$ implant, referred to as the gain layer (GL), in the case of the LGADs. 
The sensors were fabricated on wafers comprising active thicknesses of \SI{20}{\micro\meter} epitaxial layer (W3082) and a \SI{30}{\micro\meter} epitaxial layer (W3108), both grown on a \SI{300}{\micro\meter}-thick low-resistivity support wafer. 
The standard LGAD design incorporates a uniform, unpatterned p-spray layer to compensate for oxide charges and provide electrical isolation of the ($n^+$) electrodes. 
In addition, a deep junction termination extension (JTE) is implemented at the ($n^+$) electrode edges to suppress high electric fields and improve breakdown protection, as illustrated in Figure~\ref{fig:schematics}. 
The gain layer implantation energy is 380 keV, while the implantation dose is optimised to achieve the desired breakdown voltage (V$_{BD}$) \cite{GabrieleBNL}. 
All sensors feature an active area of \mbox{1.3$\times$1.3\,mm$^2$} and include a guard-ring structure surrounding the central pad. 
To enable laser illumination from the front (pad) side, the metallisation incorporates a lot\mbox{50$\times$\SI{50}{\micro\meter}$^2$} opening window. 
Photographs of the sensors investigated in this work are presented in Figure \ref{fig:schematics}, while a summary of the key parameters of the sensors is provided in Table \ref{table_1}.

\begin{table}[htbp]
\begin{center}
\caption{Electrical and geometrical parameters of the LGAD samples used in this study: full depletion voltage (V$_{FD}$), gain layer depletion voltage (V$_{GL}$), breakdown voltage (V$_{BD}$) at 20$^\circ$C, capacitance reached above full depletion (C$_{end}$) and active thickness ($d$).}
\label{table_1}
\begin{tabular}{| l | c | c | c | c | c |}
\hline
\textbf{Wafer} & \textbf{V$_{FD}$ [V]} & \textbf{V$_{GL}$ [V]} & \textbf{V$_{BD}$\,(20$^\circ$C) [V]} & \textbf{C$_{end}$}  [\si{\pico\farad}]& \textbf{\textit{d} [\si{\micro\meter}]} \\
\hline
3082 & 22  & 21  & 90 - 100  & 10.0  & 20 \\
3108    & 23  & 21  & 110 - 120  & 7.0  & 30 \\
\hline
\end{tabular}
\end{center}
\end{table}

\section{Irradiation campaign}
PINs and LGADs from each wafer were irradiated at the Proton Irradiation Facility at the Institute of Nuclear Physics (IFJ Krakow) using proton beams with energies of 30 MeV and 60 MeV. 
A hardness factor of $2.359 \pm ~ 7.5\%$ and $1.676\pm ~ 5\%$ (provided by the facility) is applied to convert the fluences to 1 MeV neutron equivalent damage for 30 MeV and 60 MeV protons, respectively. These values are in close agreement with the theoretical values calculated in \cite{summers1987correlation}.
A few PINs and LGADs were kept non-irradiated for the reference measurements. 
Following irradiation, all devices underwent a controlled annealing process for 80 minutes at \SI{60}{\degreeCelsius}. 
A summary of the irradiation fluences for the LGAD and PIN detectors is provided in Table \ref{table_2}. 
The target fluence values were intended to be the same for both proton energies. 
However, due to technical issues during the irradiation campaign, the planned fluences for the 30 MeV proton beam could not be achieved. 
As a result, a direct quantitative comparison between the two irradiation energies is not possible, and the corresponding results are therefore compared on a qualitative basis.

\begin{table}[h]
\begin{center}
\caption{Summary of the proton irradiation, including beam energy and equivalent fluence, for the LGAD and PIN detectors investigated in this work.} 
\label{table_2}
\renewcommand{\arraystretch}{1.1}
 \begin{tabular}{| c | c | c | c | c |} 
 \hline
 \textbf{Number of devices} & \textbf{Energy} & \textbf{Wafer} & \textbf{Fluence} & \textbf{Error on fluence} \\ [0.7ex] 
  &[$MeV$]  &  & [$n_{eq}/cm^2$] & [$\%$]\\
 \hline
 2 LGADs + 2 PINs & 30 & 3082 & $2.0\times10^{13}$ & $\pm$ 7.5 \\[0.5ex] 
  &  &  & $8.0\times10^{13}$ & \\[0.5ex] 
  \hline
 2 LGADs + 2 PINs & 30 & 3108 & $4.0\times10^{13}$ & $\pm$ 7.5\\[0.5ex] 
  &  &  & $6.0\times10^{13}$ & \\[0.5ex] 
  \hline
 2 LGADs + 2 PINs & 60 & 3082 & $1.0\times10^{13}$ & $\pm$ 5.0 \\[0.5ex] 
  &  &  & $1.0\times10^{14}$ & \\[0.5ex] 
  \hline
 2 LGADs + 2 PINs & 60 & 3108 & $1.0\times10^{13}$ & $\pm$ 5.0\\[0.5ex] 
  &  &  & $1.0\times10^{14}$ & \\[0.5ex] 
  \hline
\end{tabular}
\end{center}
\end{table}

\section{Electrical Characterisations}
\label{sec:elec}
\subsection{Pre-irradiation}
The electrical characterisations of all the sensors were performed using the probe station present at the AGH University of Krakow. 
The measurements were performed so that the reverse bias voltage was applied through the electrically conductive chuck. At the same time, electrical contact to the sensor pad was established using micro-positioning probe needles to measure the leakage current (I$_{leak}$). 
The current-voltage (\textit{I-V}) measurements of the non-irradiated sensors were performed with the guard ring left floating throughout the measurements. 
The capacitance-voltage (\textit{C-V}) measurements were performed using the same experimental configuration, with the addition of an LCR meter to measure the device capacitance at 1 kHz frequency. 
For non-irradiated sensors, all the measurements were performed at room temperature. 

\begin{figure}[h]
    \centering
    \begin{subfigure}[b]{0.36\textwidth}
    \centering
  \includegraphics[width=\textwidth]{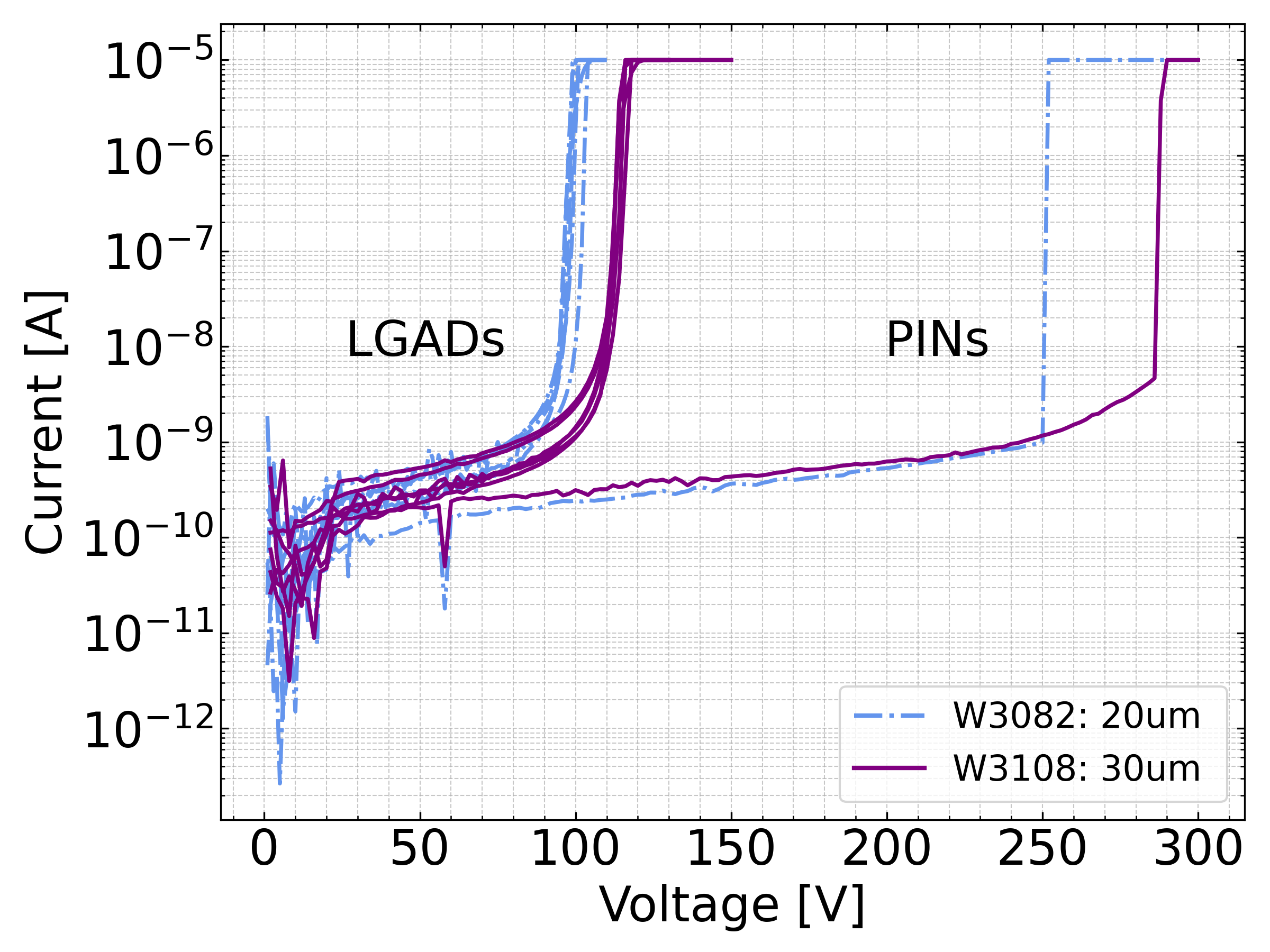}
  \caption{}
    \label{fig:iv_all}
  \end{subfigure}
      \begin{subfigure}[b]{0.36\textwidth}
        \centering
        \includegraphics[width=\textwidth]{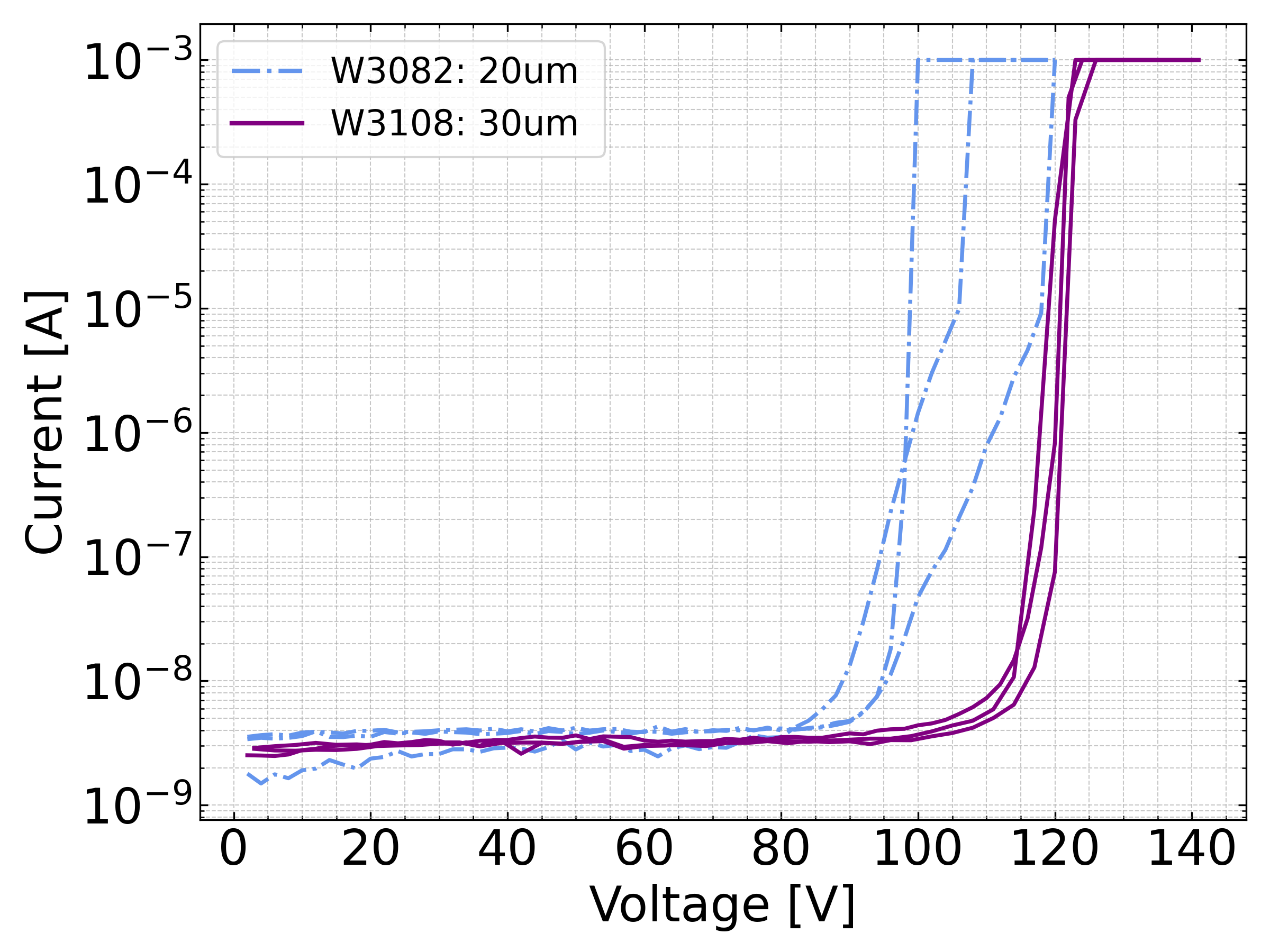}
    \caption{}
    \label{fig:iv_stress}
    \end{subfigure}
        \begin{subfigure}[b]{0.36\textwidth}
    \centering
  \includegraphics[width=\textwidth]{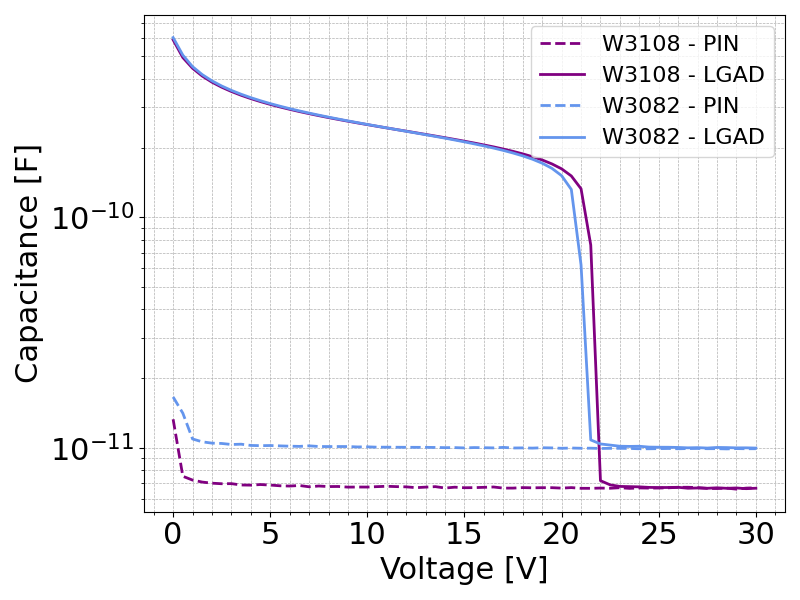}
  \caption{}
    \label{fig:cv_LGADs}
  \end{subfigure}
      \begin{subfigure}[b]{0.36\textwidth}
        \centering
        \includegraphics[width=\textwidth]{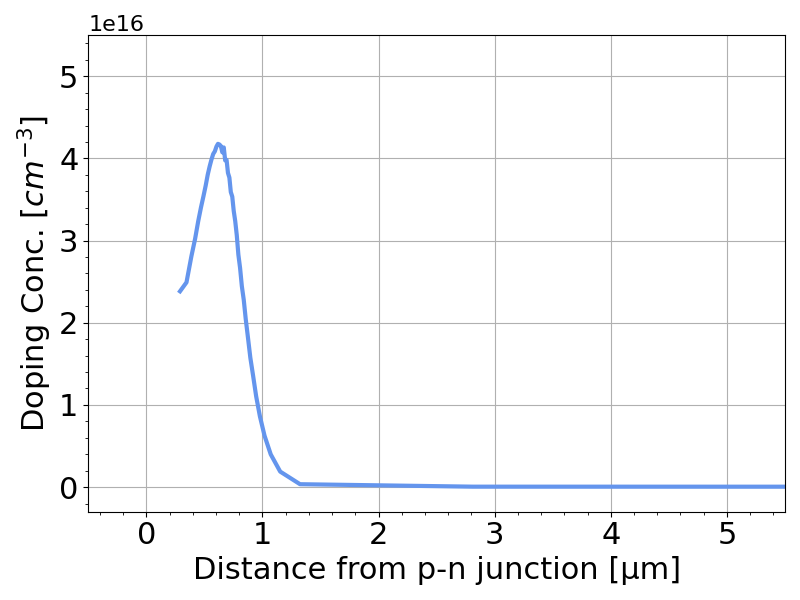}
    \caption{}
    \label{fig:dop_conc}
    \end{subfigure}
  \caption{(a) \textit{I-V} characteristics of the LGADs and PINs from W3082 and W3108. (b) \textit{I-V} taken after performing the stress test where the current compliance is set to 1 mA. (c) \textit{C-V} characteristics of the LGADs and PINs from W3082 and W3108. (d) Doping concentration obtained from the \textit{C-V} curves showing shallow gain layer implant. }
\end{figure}

Figure \ref{fig:iv_all} shows the I$_{leak}$ as a function of applied reverse bias voltage. 
The I$_{leak}$ is below nA before the V$_{BD}$, showing a well-controlled fabrication process. 
The LGADs from W3082 break down at voltages of 90-100~V, and those from W3108 break down at 110-120~V, at room temperature. 
The PIN diodes from both wafers break down in the range of 250-300~V. 
To investigate the stability of the breakdown characteristics, the LGADs were subjected to an electrical stress test in which a reverse bias above the V$_{BD}$ (V > V$_{BD}$) was applied for a fixed duration. 
After the stress period, the \textit{I-V} characteristics were remeasured using 1~mA current compliance to allow the sensors to operate well into the avalanche regime. 
The reproducibility of the \textit{I-V} curves after electrical stressing, as shown in Figure~\ref{fig:iv_stress}, shows the stability and reversibility of the breakdown behaviour.

Figure \ref{fig:cv_LGADs} shows the \textit{C-V} curves obtained for the LGADs and PINs from W3082 and W3108. 
The gain layer for both wafers depletes at the same voltage, i.e., 21 V, while full depletion occurs at 22 V and 23 V for W3082 and W3108, respectively. 
The doping concentration obtained from \textit{C-V} curves is shown in Figure~\ref{fig:dop_conc}, peaking at $\sim$ 0.6 $\mu$m from the p-n junction, exhibiting a shallow gain layer implant. 

\subsubsection{Temperature dependence on Impact Ionization }

\begin{figure}[h]
    \centering
\begin{subfigure}[b]{0.4\textwidth}
        \centering
        \includegraphics[width=\textwidth]{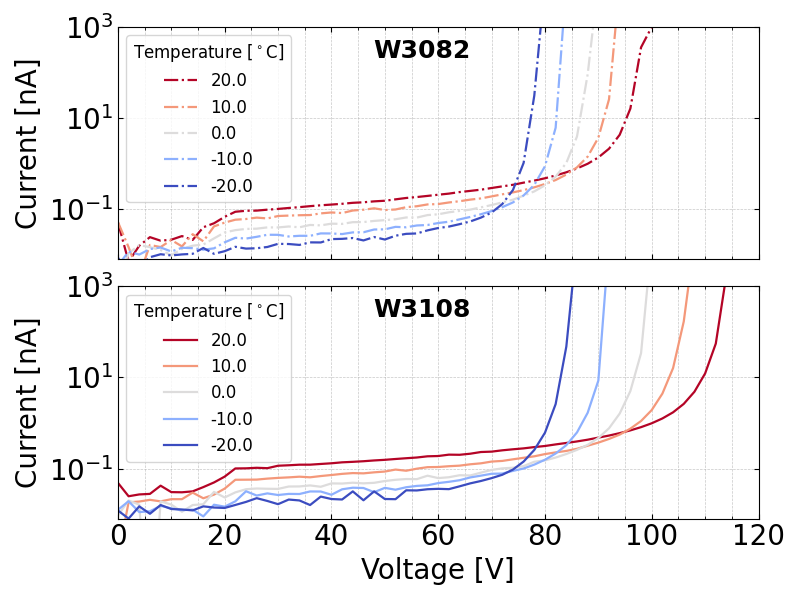}
    \caption{}
    \label{fig:iv_tempdep}
    \end{subfigure}
          \begin{subfigure}[b]{0.4\textwidth}
        \centering
        \includegraphics[width=\textwidth]{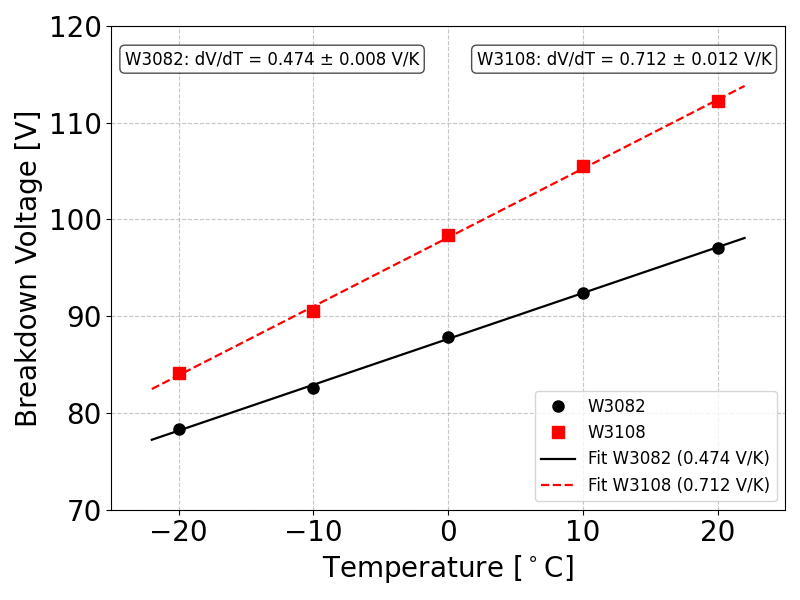}
    \caption{}
    \label{fig:vbd_temp_fit}
    \end{subfigure}
  \caption{(a) Dependence of I$_{leak}$ on varying temperature conditions. (b) Extracted V$_{BD}$ versus temperature. The linear fitting quantifies the temperature coefficient for each wafer thickness.}
\end{figure}

To investigate the influence of temperature on the avalanche characteristics, the \textit{I-V} characteristics of one non-irradiated LGAD from each wafer were measured over a temperature range from \SI{-20}{\degreeCelsius} to \SI{20}{\degreeCelsius}. 
As shown in Figure~\ref{fig:iv_tempdep}, the V$_{BD}$ increases linearly with temperature for both wafers, showing that the impact ionisation is shifted towards higher voltages at higher temperatures due to enhanced phonon scattering. 
The V$_{BD}$ was extracted from the \textit{I-V} curves where the I$_{leak}$ reaches compliance i.e., 1 $\mu$A. 

Linear fits to the extracted V$_{BD}$ yield temperature coefficients of $0.47~\mathrm{V/K}$ and $0.71~\mathrm{V/K}$ for W3082 and W3108, respectively, as shown in Figure~\ref{fig:vbd_temp_fit}. These results demonstrate an increase in the rate of V$_{BD}$ change with temperature (slope) as the active thickness increases. This relationship originates from the requirement of a higher V$_{bias}$ to initiate avalanche onset in thicker layers. Thus, the highest temperature sensitivity of V$_{BD}$ is observed for LGADs possessing the thickest active layers. These temperature coefficient values are consistent with those obtained from 20 $\mu$m thick AC-LGADs, reported in \cite{farook2026}. 

\subsection{Post-irradiation}
\begin{figure}[!h]
    \centering
    \begin{subfigure}[b]{0.36\textwidth}
    \centering
  \includegraphics[width=\textwidth]{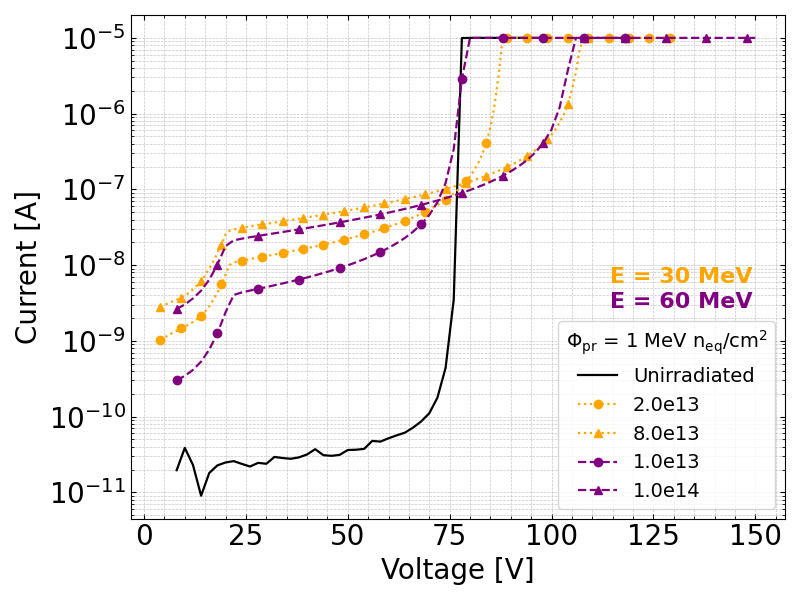}
  \caption{}
    \label{fig:iv_irr_3082}
  \end{subfigure}
      \begin{subfigure}[b]{0.36\textwidth}
        \centering
        \includegraphics[width=\textwidth]{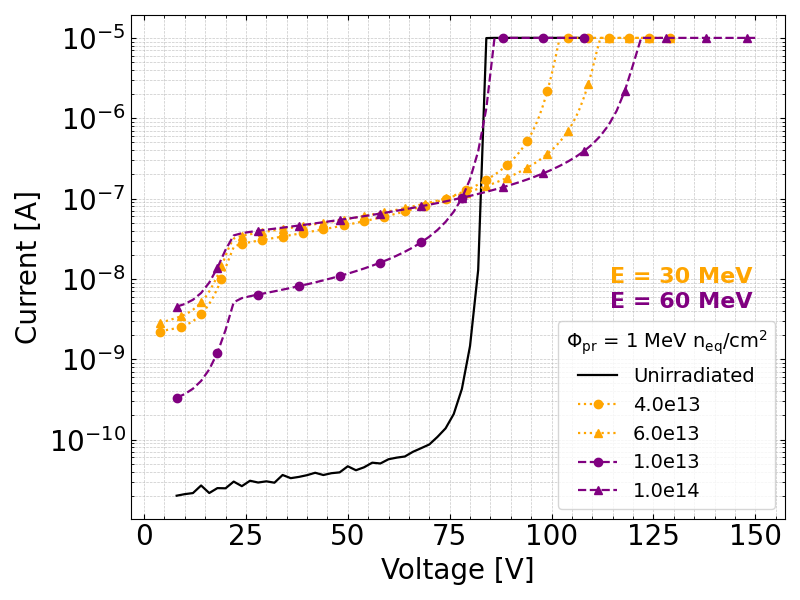}
    \caption{}
    \label{fig:iv_irr_3108}
    \end{subfigure}
        \begin{subfigure}[b]{0.36\textwidth}
    \centering
  \includegraphics[width=\textwidth]{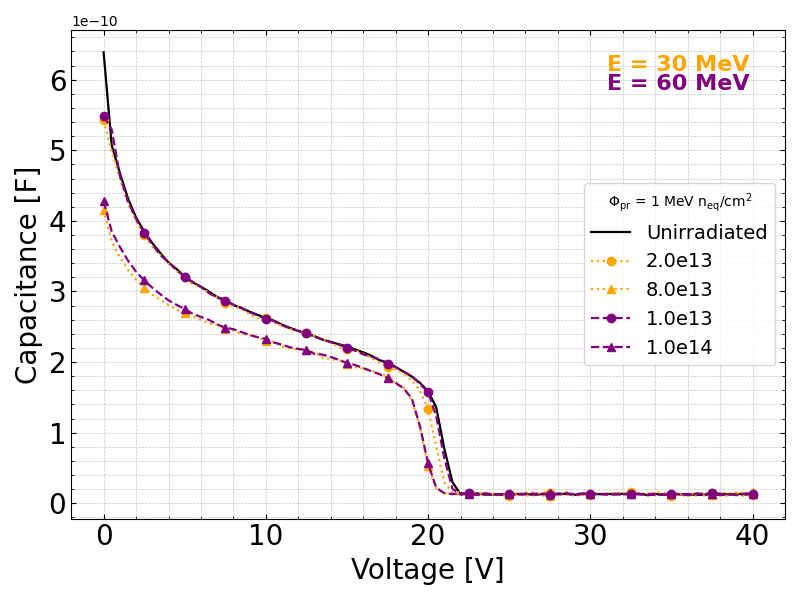}
  \caption{}
    \label{fig:cv_irr_3082}
  \end{subfigure}
      \begin{subfigure}[b]{0.36\textwidth}
        \centering
        \includegraphics[width=\textwidth]{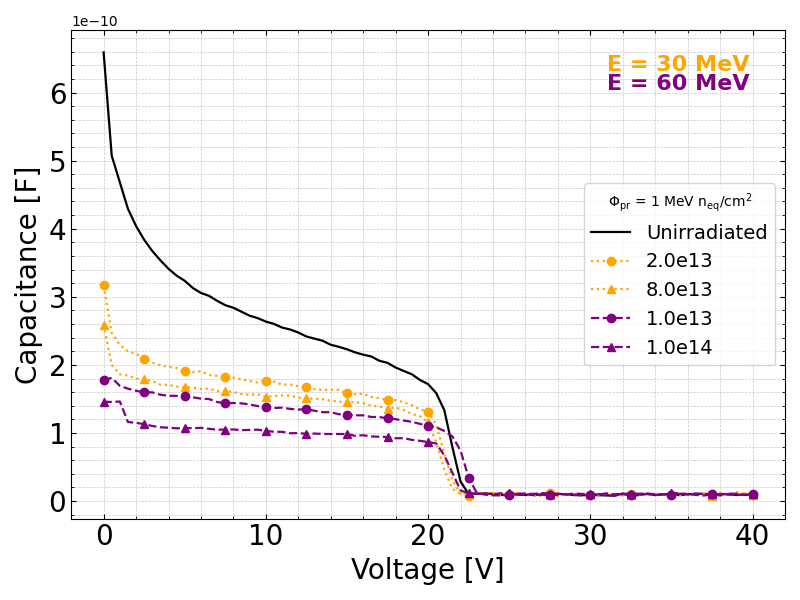}
    \caption{}
    \label{fig:cv_irr_3108}
    \end{subfigure}

  \caption{ \textit{I-V} characteristics of the LGADs after irradiation from W3082 (a) and W3108 (b). \textit{C-V} characteristics of the LGADs after irradiation from W3082 (c) and W3108 (d).}
\end{figure}

For the irradiated devices, \textit{I-V} measurements were performed at \SI{-20}{\degreeCelsius} using a climate chamber at AGH University of Krakow. 
Figures~\ref{fig:iv_irr_3082} and \ref{fig:iv_irr_3108} present the \textit{I-V} characteristics of the irradiated LGADs from W3082 and W3108, respectively. 
For both wafers, the irradiation with 30 MeV and 60 MeV protons results in a substantial increase in leakage current, referring to the creation of mid-level defects responsible for increasing I$_{leak}$. 
In the case of the 60 MeV proton beam, for both thicknesses, the lowest fluence, i.e., 1.0 $\times 10^{13}$~$n_\mathrm{eq}/cm^2$, produces only a minor increase in the V$_{BD}$, amounting to a few volts, as compared to the non-irradiated sensor. 
In contrast, irradiation to 1.0 $\times 10^{14}$~$n_\mathrm{eq}/cm^2$ shifts the V$_{BD}$ from approximately 75~V for the non-irradiated device to about 105~V for W3082 (Figure~\ref{fig:iv_irr_3082}). 
A similar trend is observed for W3108 (Figure~\ref{fig:iv_irr_3108}), where the V$_{BD}$ increases from approximately 85~V before irradiation to around 120~V after irradiating it with fluence of 1.0 $\times 10^{14}$~$n_\mathrm{eq}/cm^2$. 
These results indicate a decrease in doping concentration in the gain layer with increasing fluence. 

Although a direct comparison between the effects of the 30 MeV and 60 MeV proton irradiations is not possible because the target fluences were not identical, a qualitative comparison can still be made. 
The LGAD irradiated to 8.0 $\times 10^{13}$~n$_\mathrm{eq}/cm^2$ with 30 MeV protons shows a V$_{BD}$ similar to that of the device irradiated to 1.0 $\times 10^{14}$~n$_\mathrm{eq}/cm^2$ with 60 MeV protons, as shown in Figure~\ref{fig:iv_irr_3082}. 
This observation suggests that the lower-energy protons induce more serious bulk damage per unit fluence \cite{summers1987correlation}.

The \textit{C-V} characteristics were used to investigate the evolution of the depletion voltages following proton irradiation. 
Figures~\ref{fig:cv_irr_3082} and \ref{fig:cv_irr_3108} show the \textit{C-V} of the irradiated LGADs at different fluence levels. 
The most noticeable effect of irradiation is the gradual reduction of the V$_{GL}$ with increasing fluence. 
At a fluence of $1.0 \times 10^{13}~n_\mathrm{eq}/\mathrm{cm}^2$, only a marginal change in V$_{GL}$ is observed compared with the non-irradiated devices, indicating limited damage to the gain layer. 
In contrast, irradiation to higher fluences results in a pronounced decrease in V$_{GL}$ for both proton energies, reflecting progressive deactivation of the gain-layer acceptors due to radiation-induced defects. This reduction is consistent with the well-known acceptor removal effect in LGADs \cite{moll2020sissa, curras2023gain}.

\section{Infrared Laser Characterisations}
\label{sec:tct}
A Transient Current Technique (TCT) at AGH University of Krakow was used to perform the laser characterisations of the sensors. 
In this setup, an infrared laser (1064 nm) with a 440 ps pulse width at a repetition rate of 10 kHz is used to create the electron-hole pairs throughout the sensor bulk, mimicking a passage of a minimum ionising particle (MIP). 
The sensors are glued to an aluminium case customised for the TCT setup and mounted on a temperature-controlled, motorised three-axis stage with sub-micron precision. 
The laser was focused onto the sensor using an optical system with a spot size of $\sim$ 12 $\mu$m (FWHM). 

\subsection{Pre-Irradiation}
\begin{figure}[!ht]
    \centering
    \begin{subfigure}[b]{0.36\textwidth}
    \centering
  \includegraphics[width=\textwidth]{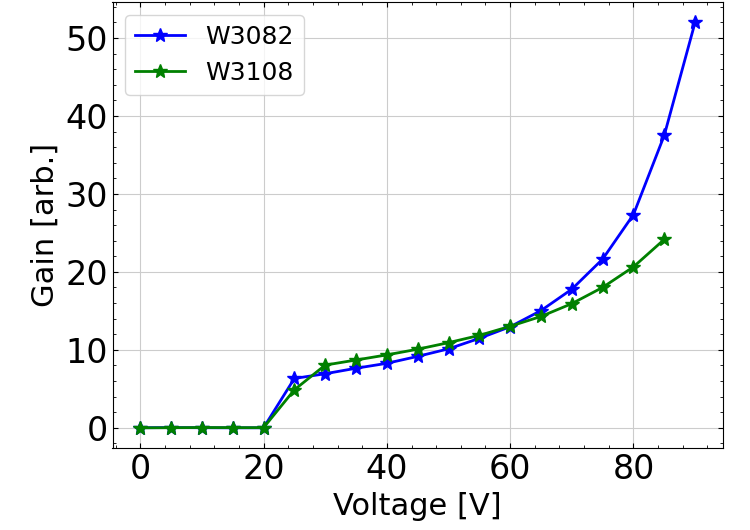}
  \caption{}
    \label{fig:Gain_3082_3108}
  \end{subfigure}
      \begin{subfigure}[b]{0.36\textwidth}
        \centering
        \includegraphics[width=\textwidth]{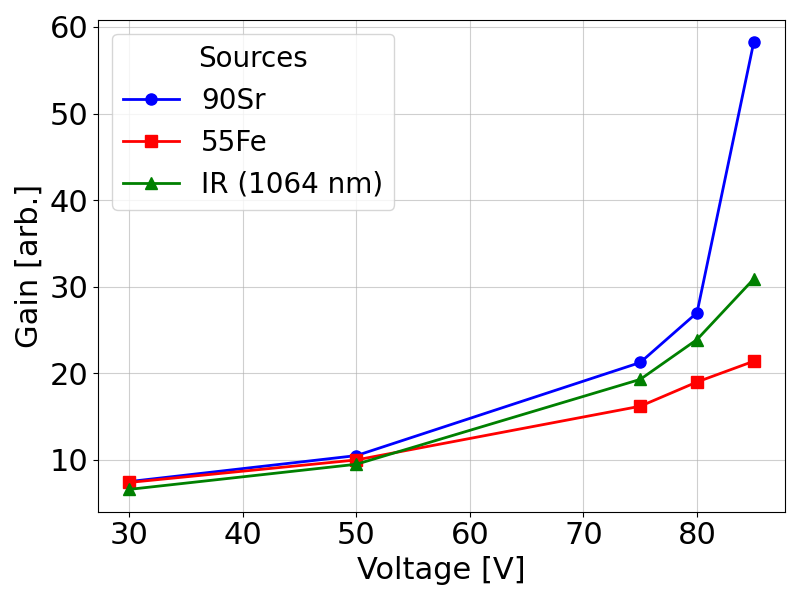}
    \caption{}
    \label{fig:gain_vs_sources}
    \end{subfigure}
  \caption{(a) Gain characteristics of the non-irradiated LGADs at room temperature from W3082 and W3108. (b) Gain of W3082 using $^{90}$Sr, $^{55}$Fe, and IR (1064 nm) laser for charge generation. }
\end{figure}

The laser intensity was calibrated to a deposited charge of $\sim$1 MIP. 
Bias voltage was applied up to 90 V in 5 V steps, remaining below the avalanche breakdown point to prevent amplitude saturation and potential amplifier damage. 
The gain is calculated using Equation \ref{eq:gain}. Up to 65 V, as both wafers exhibit comparable I-Vs, we can expect them having approximately the same gain, which, as shown in Figure \ref{fig:Gain_3082_3108}, is between 10 and 20.

\begin{equation}
    Gain = \frac{CC_{LGAD}}{CC_{PIN}(V>V_{FD})}
\label{eq:gain}
\end{equation}

Gain was further characterised using $^{90}$Sr and 
$^{55}$Fe radioactive sources at BNL, providing an independent validation of the laser-based results. The $^{90}$Sr source, with its high-energy beta particles, deposits a well-defined MIP-equivalent charge across the full sensor thickness \cite{nouicer2009status}, while the 
$^{55}$Fe source, through X-ray photon absorption, produces a localized charge deposition near the sensor surface \cite{siegmann2024development}, allowing gain to be probed under a different charge injection profile. As shown in the Figure \ref{fig:gain_vs_sources}, the gain values extracted from both source measurements are in good agreement with the IR laser measurements performed at AGH University of Krakow, confirming the consistency of the gain characterisation across independent measurement techniques and facilities.

\subsection{Post-Irradiation}

\begin{figure}[!ht]
    \centering
    \begin{subfigure}[b]{0.36\textwidth}
    \centering
  \includegraphics[width=\textwidth]{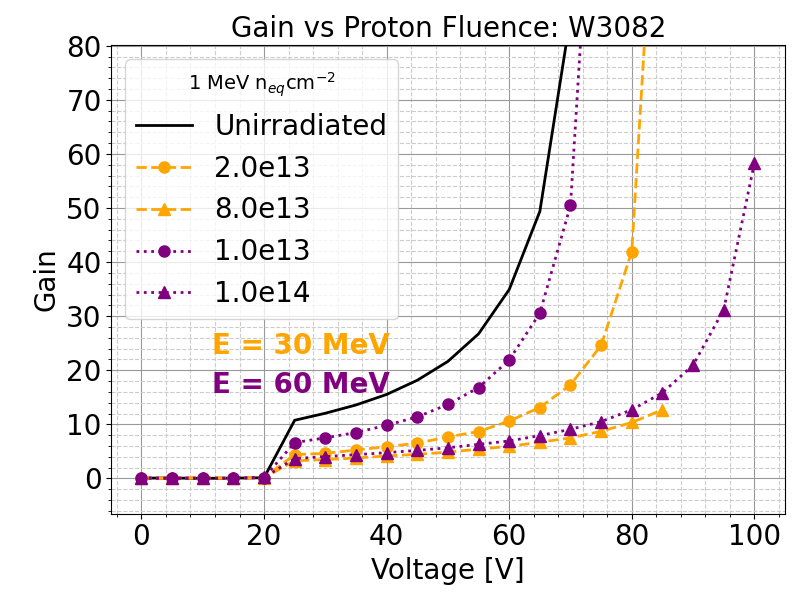}
  \caption{}
    \label{fig:gain_irr_3082}
  \end{subfigure}
      \begin{subfigure}[b]{0.36\textwidth}
        \centering
        \includegraphics[width=\textwidth]{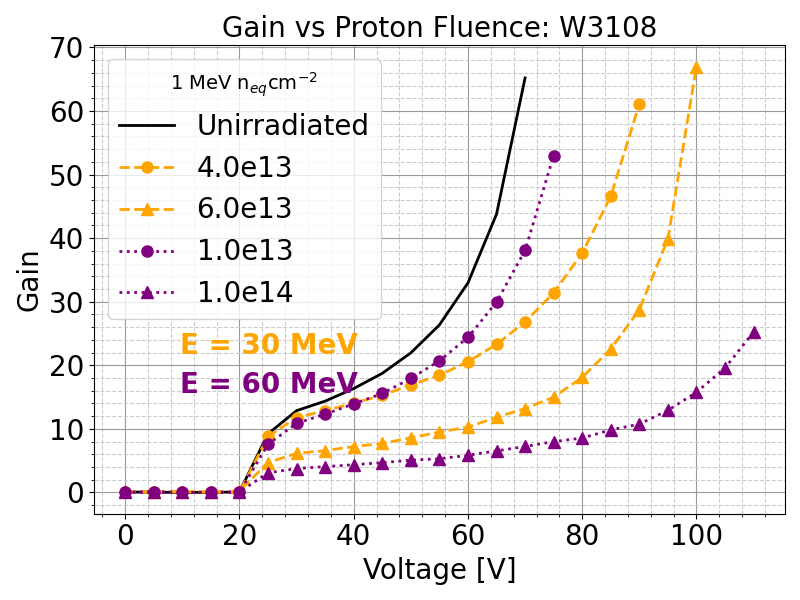}
    \caption{}
    \label{fig:gain_irr_3108}
    \end{subfigure}
  \caption{ Gain characteristics of the LGADs before and after irradiation from (a) W3082 and (b) W3108, at \SI{-20}{\degreeCelsius}.}
\end{figure}

Figures \ref{fig:gain_irr_3082} and \ref{fig:gain_irr_3108} show the gain as a function of bias voltage for irradiated LGADs from wafers W3082 and W3108, compared to non-irradiated reference sensors, following irradiation with 30 MeV and 60 MeV protons at various 1 MeV n$_{eq}$cm$^{-2}$ fluences given in Table \ref{table_2}. For both wafers, the non-irradiated devices exhibit a steep rise in gain with voltage, reaching breakdown at relatively low bias ($\sim$70–75 V). After irradiation, the gain curves are significantly suppressed and avalanche breakdown is shifted to higher voltages, which reduces the effective doping concentration of the gain layer and, consequently, the electric field available for impact ionisation. As a result, higher bias voltages are required to reach a given gain value, and the V$_{BD}$ increases progressively with fluence.

This trend is consistent across both energies and fluences studied: at a fixed voltage, devices irradiated to lower fluences (e.g., $1.0\times10^{13}$ n$_{\rm eq}$/cm$^2$) retain higher gain than those irradiated to higher fluences (e.g., $1.0\times10^{14}$ n$_{\rm eq}$/cm$^2$), confirming that gain degradation scales with fluence. Notably, even at the highest fluence of $1.0\times10^{14}$ n$_{\rm eq}$/cm$^2$, a gain above 10 is still observed for both wafers, indicating that despite substantial radiation damage to the gain layer, the devices retain sufficient multiplication to provide an adequate signal-to-noise ratio (SNR) for particle detection, which is essential for maintaining timing and tracking performance in EIC.

\section{Timing Resolution}
\label{sec:timing}
Timing resolution measurements were performed using a TCT setup with a laser-splitting configuration, in which the IR laser pulse was split into two paths with a fixed delay of 120 ns between them, allowing simultaneous triggering and timing comparison between the two signal branches. 
By this method, the timing jitter is extracted by taking the time difference between the two split pulses, fitting the resulting distribution with a Gaussian, and dividing the standard deviation of the fit by $\sqrt{2}$. 
To eliminate time-walk effects, the arrival time of each pulse is determined using a constant fraction discrimination (CFD) method, with the threshold set at 30\% of the pulse amplitude. 
Signals were digitised using an Agilent Infinium DSA 80904A oscilloscope with an 8 GHz bandwidth and a sampling rate of 10 GSa/s. 
To ensure a sufficiently high SNR for precise timing extraction, the laser intensity was set to deposit a charge equivalent to $\sim$5 MIPs, minimising the impact of electronic noise on the time-of-arrival measurement.

\begin{figure}[!ht]
    \centering
    \begin{subfigure}[b]{0.38\textwidth}
    \centering
  \includegraphics[width=\textwidth]{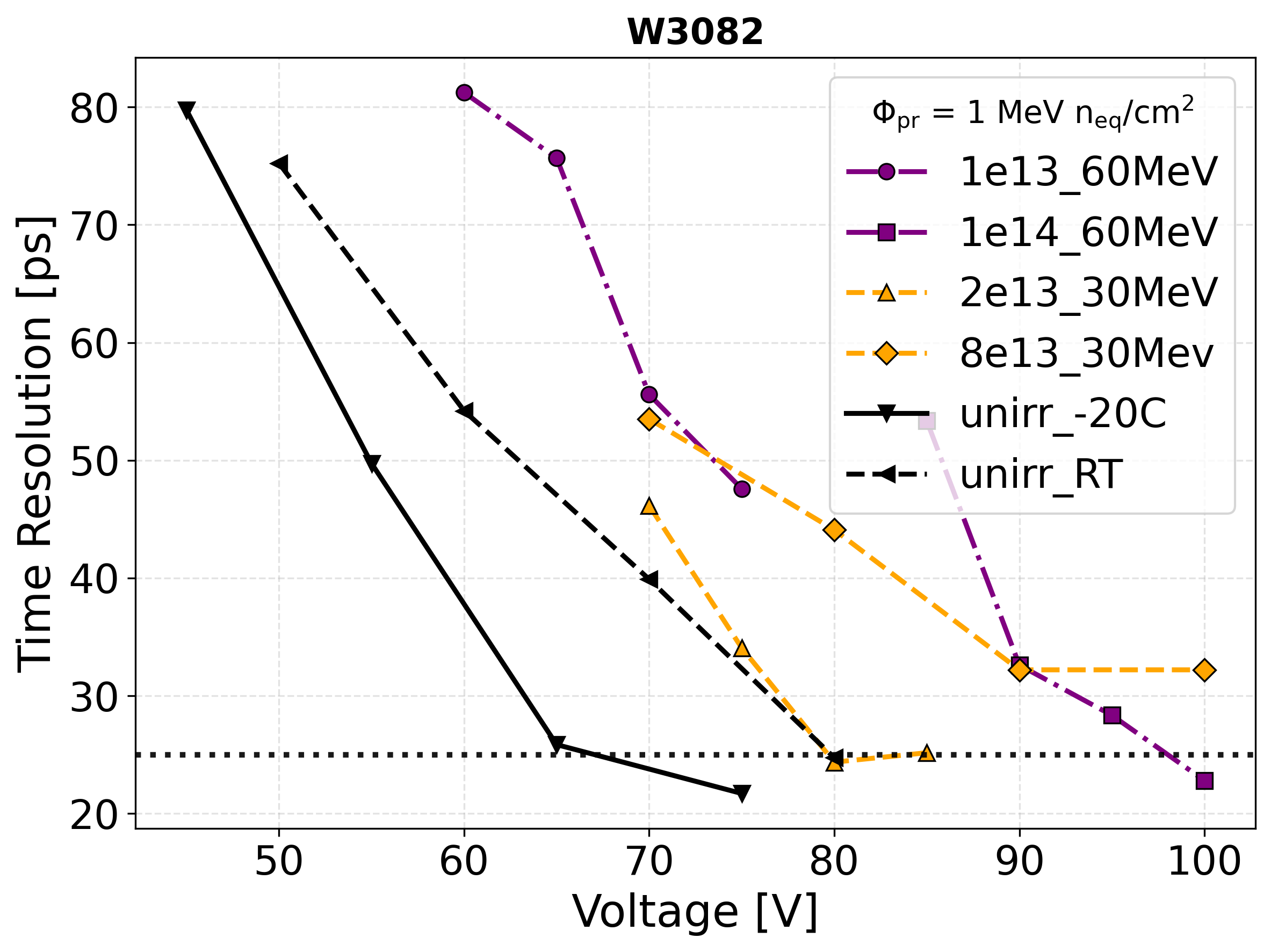}
  \caption{}
    \label{fig:timing_W3082}
  \end{subfigure}
      \begin{subfigure}[b]{0.38\textwidth}
        \centering
        \includegraphics[width=\textwidth]{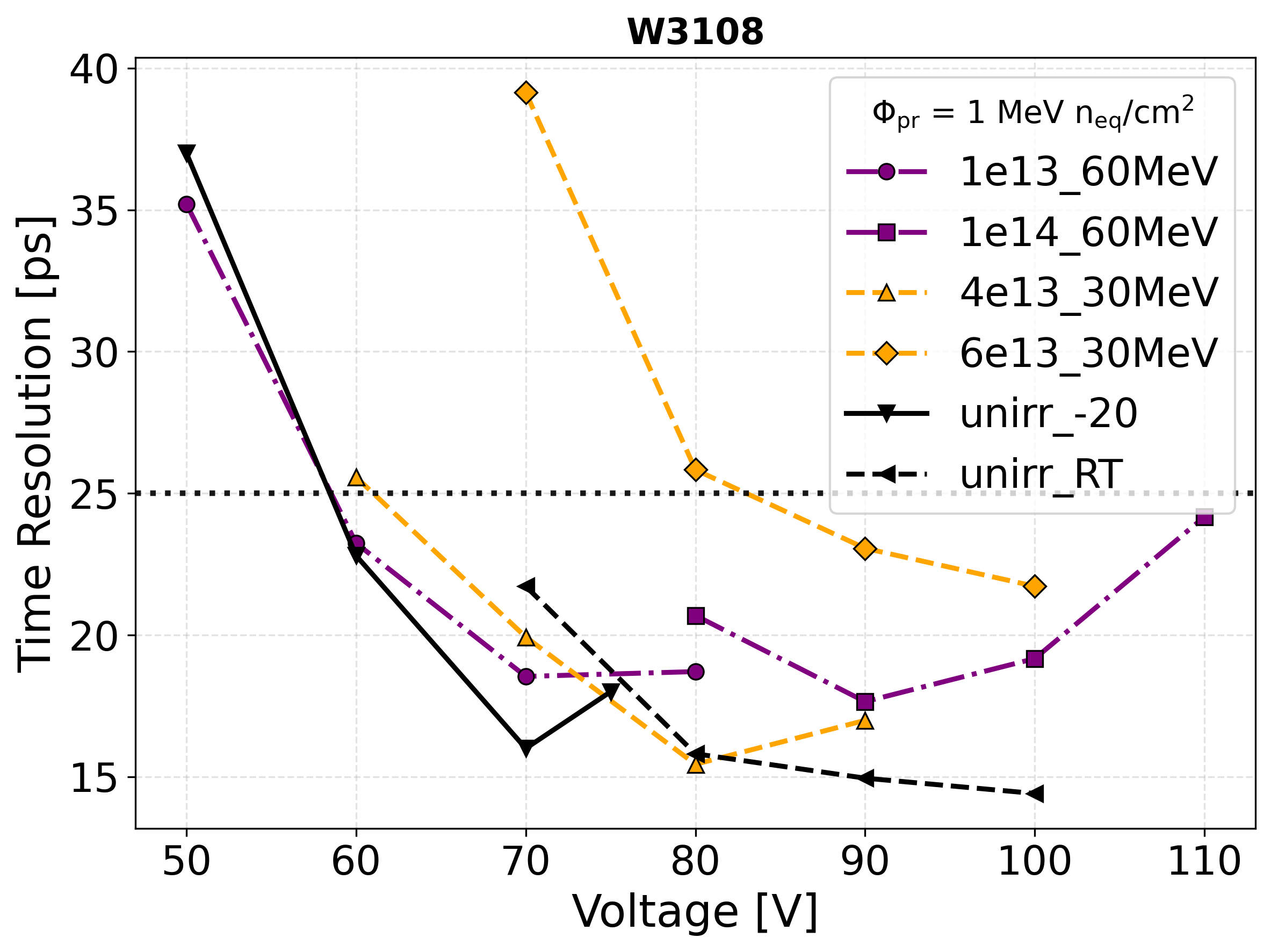}
    \caption{}
    \label{fig:timing_W3108}
    \end{subfigure}
  \caption{A comparison of the timing resolution as a function of voltage of non-irradiated and irradiated devices from (a) W3082 and (b) W3108. The dotted line shows a baseline of 25 ps for reference.}
\end{figure}

Figures \ref{fig:timing_W3082} and \ref{fig:timing_W3108} show the time resolution as a function of bias voltage for non-irradiated and irradiated LGADs from wafers W3082 and W3108. Non-irradiated devices were measured both at room temperature and at \SI{-20}{\degreeCelsius}, while irradiated devices, exposed to 30 MeV and 60 MeV protons at fluences ranging from $1.0\times10^{13}$ to $1.0\times10^{14}$ n$_{\rm eq}$/cm$^2$, were measured at \SI{-20}{\degreeCelsius}. 

For both wafers, the time resolution improves with increasing bias voltage, as expected from the corresponding increase in gain and signal slew rate. Non-irradiated devices achieve the best time resolution at relatively low voltages, reaching below 25 ps around 65–75 V for W3082 and 55–60 V for W3108. For non-irradiated devices, the time resolution is obtained at room temperature and \SI{-20}{\degreeCelsius}, showing that timing resolution at the same voltage improves at lower temperatures for both wafer thicknesses. Irradiated devices require substantially higher bias voltages to reach comparable time resolution, consistent with the radiation-induced reduction in gain discussed previously; this shift is more pronounced for the higher-fluence samples ($1.0\times10^{14}$ n$_{\rm eq}$/cm$^2$), which need voltages close to or above 90–100 V to reach timing resolution below 25 ps, which is comparable to the results reported in thin-LGADs from other groups \cite{giacomini2019fabrication, SUN2024169203, WHITE2026171596}. It should be noted that Landau fluctuations are absent in the laser-based timing resolution setup. These results should be regarded as preliminary, as the timing resolution was limited by high capacitance caused by the long cables connecting the sensor to the amplifier, and no dedicated fast-timing boards were used to mitigate this effect. Timing resolution can be improved with fast-timing compatible readout electronics.

\section{Summary and Conclusions}
\label{sec:summary}
This work presents a systematic study of LGADs fabricated at BNL on 20 $\mu$m and 30 $\mu$m epitaxial substrates, characterised before and after irradiation with 30 MeV and 60 MeV protons at fluences from $1.0\times10^{13}$ up to $1.0\times10^{14}$ n$_{\rm eq}$/cm$^2$. The study focused on electrical characterisation, gain, and timing performance. Before irradiation, \textit{I-V} and \textit{C-V} measurements showed well-formed gain layers, leakage currents below the nA level before breakdown, and a reproducible breakdown behaviour after electrical stressing. The V$_{BD}$ increased linearly with temperature, with coefficients of 0.47 V/K and 0.71 V/K for the 20 $\mu$m and 30 $\mu$m sensors. Gain measurements with an IR laser agreed with $^{90}$Sr and $^{55}$Fe source measurements, with both wafers reaching a gain of about 15 at 65~V. Before irradiation, a time resolution of 22~ps for W3082 and 15~ps for W3108 was achieved. 

After irradiation, the \textit{I-V}, \textit{C-V}, gain and timing characteristics changed with fluence. The leakage current increased, V$_{BD}$ shifted to higher values, and the V$_{GL}$ decreased. This is consistent with deactivation of gain layer acceptors through the acceptor removal effect. The gain was suppressed at a given bias, but it remained above 5 even at the highest fluence, giving enough signal-to-noise ratio for particle detection. Since the target fluences for the two proton energies were not identical, they could only be compared qualitatively. The gain comparison suggests that 30 MeV protons cause more bulk damage per unit fluence than 60 MeV protons. Irradiated devices recovered the time resolution of < 25~ps for W3082 and 15~ps for W3108 once sufficient bias was applied. 

\section*{Acknowledgments}
This work was supported by  \textit{Excellence Initiative – Research University AGH IDUB 6275} and the \textit{National Science Centre, Poland (NCN), UMO-2018/31/B/ST2/03998} grants, and received funding from the European Union’s Horizon Europe Research and Innovation programme under Grant Agreement No 101057511 (EURO-LABS). 



\bibliographystyle{elsarticle-num}
\bibliography{bib}

@article{Sadrozinski_2018,
doi = {10.1088/1361-6633/aa94d3},
year = {2017},
month = {dec},
publisher = {IOP Publishing},
volume = {81},
number = {2},
pages = {026101},
author = {Sadrozinski, Hartmut F-W and Seiden, Abraham and Cartiglia, Nicolò},
title = {4D tracking with ultra-fast silicon detectors},
journal = {Reports on Progress in Physics}
}

@article{SADROZINSKI201618,
title = {Ultra-{f}ast {s}ilicon {d}etectors ({UFSD})},
journal = {Nuc. Inst. and Meth. in Phys. Res. Sec. A: Acc., Spec., Det. and Ass. Equip.},
volume = {831},
pages = {18-23},
year = {2016},
note = {},
issn = {0168-9002},
doi = {10.1016/j.nima.2016.03.093},
url = {},
author = {H.F.-W. Sadrozinski et. al.},

}

@article{pellegrini2014technology,
  title={Technology developments and first measurements of {L}ow {G}ain {A}valanche {D}etectors ({LGAD}) for high energy physics applications},
  author={Pellegrini, Giulio and Fern{\'a}ndez-Mart{\'\i}nez, Pablo and Baselga, Marta and Fleta, Celeste and Flores, D and Greco, Virginia and Hidalgo, Salvador and Mandi{\'c}, I and Kramberger, G and Quirion, David and others},
  journal={Nuclear Instruments and Methods in Physics Research Section A: Accelerators, Spectrometers, Detectors and Associated Equipment},
  volume={765},
  pages={12--16},
  year={2014},
  publisher={Elsevier},
  
  doi = "10.1016/j.nima.2014.06.008"
}

@article{sola2025thin,
  title={Thin {LGAD} sensors for {4D} tracking in high radiation environments: state of the art and perspectives},
  author={Sola, Valentina and Boscardin, Maurizio and Moscatelli, Francesco and Altamura, Anna Rita and Arcidiacono, Roberta and Borghi, Giacomo and Cartiglia, Nicolo and Centis Vignali, Matteo and Croci, Tommaso and Durando, Matteo and others},
  journal={Frontiers in Sensors},
  volume={6},
  pages={1648102},
  year={2025},
  doi={10.3389/fsens.2025.1648102},
  publisher={Frontiers Media SA}
}

@article{curras2020inverse,
  title={Inverse {L}ow {G}ain {A}valanche {D}etectors (i{LGAD}s) for precise tracking and timing applications},
  author={Curr{\'a}s, Esteban and Carulla, M and Vignali, M Centis and Duarte-Campderros, J and Fern{\'a}ndez, M and Flores, David and Garc{\'\i}a, A and G{\'o}mez, G and Gonz{\'a}lez, J and Hidalgo, Salvador and others},
  journal={Nuclear Instruments and Methods in Physics Research Section A: Accelerators, Spectrometers, Detectors and Associated Equipment},
  volume={958},
  pages={162545},
  year={2020},
  publisher={Elsevier},

  doi="10.1016/j.nima.2019.162545"
}

@article{giacomini2019fabrication,
  title={Fabrication and performance of {AC}-coupled {LGAD}s},
  author={Giacomini, Gabriele and Chen, Wei and D'Amen, Gabriele and Tricoli, Alessandro},
  journal={Journal of Instrumentation},
  volume={14},
  number={09},
  pages={P09004},
  year={2019},
  publisher={IOP Publishing},

  doi="10.1088/1748-0221/14/09/P09004"
}

@article{paternoster2020trench,
  title={Trench-isolated low gain avalanche diodes (ti-{LGAD}s)},
  author={Paternoster, G and Borghi, G and Boscardin, Maurizio and Cartiglia, Nicolo and Ferrero, Maricel and Ficorella, F and Siviero, Federico and Gola, Alberto and Bellutti, Pierluigi},
  journal={IEEE Electron Device Letters},
  volume={41},
  number={6},
  pages={884--887},
  year={2020},
  publisher={IEEE},

  doi="10.1109/LED.2020.2991351"
}

@article{Zareef_2025,
doi = {10.1088/1748-0221/20/07/P07037},

year = {2025},
month = {jul},
publisher = {IOP Publishing},
volume = {20},
number = {07},
pages = {P07037},
author = {Zareef, Fasih and Moffat, Neil and Bates, Richard and Maneuski, Dzmitry and Han, Ruirui and Bullough, Mark and Williams, Mark and Obłąkowska-Mucha, Agnieszka and Szumlak, Tomasz},
title = {{Achieving near 100\% Fill factor for small pixel LGADs: Study of First Trench Isolated LGADs fabricated at Micron Semiconductor Ltd}},
journal = {Journal of Instrumentation},

}

@article{yano2025physics,
  title={{Physics Opportunity at the Electron-Ion Collider (EIC)}},
  author={Yano, Satoshi},
  journal={arXiv preprint arXiv:2505.05780},
  doi={10.48550/arXiv.2505.05780},
  year={2025}
}

@article{STAGE2026171012,
title = {{Performance of neutron and proton irradiated AC-LGAD sensors}},
journal = {Nuclear Instruments and Methods in Physics Research Section A: Accelerators, Spectrometers, Detectors and Associated Equipment},
volume = {1082},
pages = {171012},
year = {2026},
issn = {0168-9002},
doi = {10.1016/j.nima.2025.171012},

author = {G. Stage and A. Borjigin and J. Ding and M. Davis and S. Beringer and M. Gignac and F. McKinney-Martinez and S.M. Mazza and A. Molnar and J. Ott and H.F.-W. Sadrozinski and B. Schumm and A. Seiden and T. Shin and M. Wilder and G. Kramberger and I. Mandic and S. Seidel and J. Si and R. Novotny}
}

@article{mazza2019high,
  title={{A High-Granularity Timing Detector (HGTD) for the Phase-II upgrade of the ATLAS detector}},
  author={Mazza, Simone Michele},
  journal={Journal of Instrumentation},
  volume={14},
  number={10},
  doi={10.1088/1748-0221/14/10/C10028},
  pages={C10028--C10028},
  year={2019}
}

@article{ferrero2022cms,
  title={{The CMS MTD Endcap Timing Layer: Precision timing with Low Gain Avalanche Diodes}},
  author={Ferrero, Marco and CMS Collaboration and others},
  journal={Nuclear Instruments and Methods in Physics Research Section A: Accelerators, Spectrometers, Detectors and Associated Equipment},
  doi={10.1016/j.nima.2022.166627},
  volume={1032},
  pages={166627},
  year={2022},
  publisher={Elsevier}
}

@misc{GabrieleBNL,
author={Giacomini,Gabriele and Boukhicha, M and Tsang, T},
Title={{Optical Characterization of
Low-Gain Avalanche Diodes}},

doi = {10.1109/NSS/MIC/RTSD57106.2025.11287424}
}

@article{summers1987correlation,
  title={Correlation of particle-induced displacement damage in silicon},
  author={Summers, Geoffrey P and Burke, EA and Dale, CJ and Wolicki, EA and Marshall, PW and Gehlhausen, MA},
  doi={10.1109/TNS.1987.4337442},
  journal={IEEE Transactions on Nuclear Science},
  volume={34},
  number={6},
  pages={1133--1139},
  year={1987},
  publisher={IEEE}
}

@article{moll2020sissa,
  title={{Acceptor removal-Displacement damage effects involving the shallow acceptor doping of p-type silicon devices}},
  author={Moll, Michael},
  journal={PoS},
  pages={027},
  year={2020},
  doi={10.22323/1.373.0027},
  publisher={SISSA},
}

@article{curras2023gain,
  title={{Gain layer degradation study after neutron and proton irradiations in Low Gain Avalanche Diodes}},
  author={Curras Rivera, Esteban and La Rosa, Alessandro and Moll, Michael and Zareef, Fasih},
  journal={Journal of Instrumentation},
  volume={18},
  number={10},
  pages={P10020},
  year={2023},
  doi={10.1088/1748-0221/18/10/P10020},
  publisher={IOP Publishing}
}

@article{nouicer2009status,
  title={{Status and performance of new silicon stripixel detector for the PHENIX experiment at RHIC: Beta source, cosmic-rays and proton beam at 120 GeV}},
  author={Nouicer, Rachid and Akiba, Yasuyuki and Bennett, Robert and Boyle, Kieran and Cianciolo, Vince and Deshpande, Abhay and Dion, Alan and Eggleston, Michael and Enokizono, Akitomo and Kaneti, Steve and others},
  journal={Journal of Instrumentation},
  volume={4},
  number={04},
  pages={P04011--P04011},
  doi={10.1088/1748-0221/4/04/P04011},
  year={2009}
}

@article{siegmann2024development,
  title={{Development of a silicon drift detector array to search for keV-scale sterile neutrinos with the KATRIN experiment}},
  author={Siegmann, Daniel and Edzards, Frank and Bruch, Christina and Biassoni, Matteo and Carminati, Marco and Descher, Martin and Fiorini, Carlo and Forstner, Christian and Gavin, Andrew and Gugiatti, Matteo and others},
  journal={Journal of Physics G: Nuclear and Particle Physics},
  volume={51},
  number={8},
  pages={085202},
  year={2024},
  doi={ 10.1088/1361-6471/ad4bf8},
  publisher={IOP Publishing}
}

@misc{farook2026,
      title={{Impact of Environmental Stress on Low Gain Avalanche Diode Sensors Response}}, 
      author={Mohamed Hijas Mohamed Farook and Gaetano Barone and Diallo Boye and Gabriele D'Amen and Gabriele Giacomini and Timo Peltola and Jennifer Roloff and Enrico Rossi and Trevor Russell and Sally Seidel and Alessandro Tricoli and Lixing Wang},
      year={2026},
      primaryClass={physics.ins-det},
      doi={10.48550/arXiv.2608.22482}
       
}

@article{zareef2026depth,
  title={Depth-dependent inter-pixel distance in trench-isolated LGADs for shallow radiation detection},
  author={Zareef, Fasih and Mianroodi, Parisa Rezaei and Rebarz, Mateusz and Fernandez-Mart{\'\i}nez, Pablo and Bates, Richard and Maneuski, Dzmitry and Bullough, Mark and Ob{\l}akowska-Mucha, Agnieszka and Szumlak, Tomasz and Moffat, Neil},
  journal={Scientific Reports},
  year={2026},
  publisher={Nature Publishing Group UK, London},
  doi={10.1038/s41598-026-74942-1}
}

@article{KRAUS2026171623,
title = {Proton energy dependence of radiation induced Low Gain Avalanche Detector degradation},
journal = {Nuclear Instruments and Methods in Physics Research Section A: Accelerators, Spectrometers, Detectors and Associated Equipment},
volume = {1090},
pages = {171623},
year = {2026},
issn = {0168-9002},
doi = {10.1016/j.nima.2026.171623},
author = {Veronika Kraus and Marcos Fernández García and Luca Menzio and Michael Moll},

}

@article{TISHELMANCHARNY2027172107,
title = {Mortality of ultra-thin LGADs and PiN diodes from high energy deposition},
journal = {Nuclear Instruments and Methods in Physics Research Section A: Accelerators, Spectrometers, Detectors and Associated Equipment},
volume = {1094},
pages = {172107},
year = {2027},
issn = {0168-9002},
doi = {10.1016/j.nima.2026.172107},
author = {A. Tishelman-Charny and A. Buzzi and F. Capocasa and G. D’Amen and S. Diaw and D. Duan and M.H. Mohamed Farook and G. Giacomini and M. Kurth and D. Ponman and J. Roloff and E. Rossi and S. Stucci and A. Tricoli and H. Zhang},
}

@article{SUN2024169203,
title = {The performance of AC-coupled Strip LGAD developed by IHEP},
journal = {Nuclear Instruments and Methods in Physics Research Section A: Accelerators, Spectrometers, Detectors and Associated Equipment},
volume = {1062},
pages = {169203},
year = {2024},
issn = {0168-9002},
doi = {10.1016/j.nima.2024.169203},
author = {Weiyi Sun and Mengzhao Li and Zhijun Liang and Mei Zhao and Xiaoxu Zhang and Tianyuan Zhang and Yuan Feng and Shuqi Li and Xinhui Huang and Yunyun Fan and Tianya Wu and Xuan Yang and Bo Liu and Wei Wang and Yuekun Heng and Gaobo Xu and João Guimaraes {da Costa}},
}

@article{WHITE2026171596,
title = {Timing resolution from beam tests on thin LGADs down to 16.6 ps},
journal = {Nuclear Instruments and Methods in Physics Research Section A: Accelerators, Spectrometers, Detectors and Associated Equipment},
volume = {1089},
pages = {171596},
year = {2026},
issn = {0168-9002},
doi = {10.1016/j.nima.2026.171596},
author = {Robert Stephen White and Marco Ferrero and Valentina Sola and Anna Rita Altamura and Roberta Arcidiacono and Maurizio Boscardin and Nicolò Cartiglia and Matteo Centis Vignali and Tommaso Croci and Matteo Durando and Simone Galletto and Alessandro Fondacci and Leonardo Lanteri and Ludovico Massaccesi and Luca Menzio and Arianna Morozzi and Francesco Moscatelli and Daniele Passeri and Giovanni Paternoster and Federico Siviero},
}

\end{document}